\documentclass[fleqn,usenatbib]{mnras}

\usepackage{newtxtext,newtxmath}

\usepackage[T1]{fontenc}

\DeclareRobustCommand{\VAN}[3]{#2}
\let\VANthebibliography\thebibliography
\def\thebibliography{\DeclareRobustCommand{\VAN}[3]{##3}\VANthebibliography}

\usepackage{graphicx}	
\usepackage{amsmath}	
\usepackage{threeparttable}
\usepackage{subcaption}
\usepackage{pdflscape}
\defcitealias{lomaeva_22}{Paper I}

\newcommand{\dsfr}{$\langle  SFR_{\textnormal{5}} \rangle \big/  \langle SFR_{\textnormal{200}} \rangle$}

\title[The recent SFHs of nearby galaxies]{The recent star formation histories of nearby galaxies on resolved scales}

\author[M. Lomaeva et al.]{
Maria Lomaeva,$^{1}$\thanks{E-mail:maria.lomaeva.19@ucl.ac.uk}
Amélie Saintonge,$^{1}$
Ilse De Looze,$^{1,2}$
\\
$^{1}$Dept. of Physics \& Astronomy, University College London, Gower Street, London WC1E 6BT, UK\\
$^{2}$Sterrenkundig Observatorium, Ghent University, Krijgslaan 281 - S9, 9000 Gent, Belgium\\
}

\date{Accepted XXX. Received YYY; in original form ZZZ}

\pubyear{2023}

\begin{document}
\label{firstpage}
\pagerange{\pageref{firstpage}--\pageref{lastpage}}
\maketitle

\begin{abstract}
Star formation histories (SFHs) of galaxies are affected by a variety of factors, both external (field vs. cluster/group) and internal (presence of a bar and AGN, morphological type). In this work, we extend our previous study and apply the \dsfr\, metric to a sample of eleven nearby galaxies with MUSE observations. Based on a combination of H$\alpha$ and UV photometry, \dsfr\ is sensitive to star formation timescales of $\sim$5--200~Myr and therefore measures the present-day rate of change in the star formation rate, dSFR/dt. Within this limited galaxy sample, we do not observe systematic variations between the global value of \dsfr\ and the presence of an active galactic nucleus, stellar bar, nor with group or cluster membership. Within some of the individual galaxies, we however observe significant differences in \dsfr\ between the arm and interarm regions. In half of the galaxies, the recent SFH of both arm and interarm regions has been very similar.  However, in the galaxies with higher bulge-to-total light ratios and earlier morphological type, the SFR is declining more rapidly in the interarm regions. This decline in SFR is not a result of low molecular gas surface density or a decrease in the star formation efficiency, implying that other factors are responsible for this SFR decrease.  

\end{abstract}

\begin{keywords}
galaxies: spiral -- galaxies: star formation -- galaxies: evolution
\end{keywords}



\section{Introduction}

Star formation (SF) changes in galaxies can occur both on global and local physical scales over a variety of time periods. Global and long-term changes are responsible for the blue and red sequence of galaxies illustrated by the bimodal distribution of galaxies in the colour-magnitude diagram or the star-forming main sequence (SFMS) \citep[e.g.][]{noeske_07, Wuyts_11, van_der_Wel_14}. Non-permanent as well as local changes inside a galaxy are responsible for the oscillations and scatter around the SFMS \citep[e.g.][]{tacchella16}. 

A galaxy's ability to form stars is constrained by the availability of cold and dense gas, with a range of processes that can affect gas reservoirs and their efficiency at forming stars. In rich galaxy environments, such as groups and clusters, a galaxy can become significantly depleted of gas through ram-pressure stripping \citep[e.g.,][]{gunn_72, Abadi_99, Quilis_00,Boselli_22} or strangulation, when further gas accretion is inhibited by the removal, heating or stabilisation of the gaseous envelope around a galaxy \citep{larson_74, Balogh_00, Balogh_2000, keres_05,Dekel_06}. However, if a galaxy manages to resume gas accretion, its SF activity may become rejuvenated, even if it was previously terminated or heavily subdued \citep{Lemonias_11, Fang_12}.

AGN activity has long been known to decrease or quench SF via AGN-driven outflows that remove gas from the galaxy and suppress the cold accretion \citep[e.g.,][]{Croton_06,Heckman_14}. However, if the power output of the AGN reduces, the SF can reignite gain. Moreover, AGN jets can conversely shock the gas and compress the ISM, creating cocoons of turbulent gas, where molecular hydrogen is able to clump more efficiently and form stars \citep[e.g.,][]{Silk_10}. In simulations, a subsequent increase in the SF can also occur further out in the disk due to the pressurised ISM \citep{Gaibler_12, Ishibashi_12}. These opposing effects of AGN feedback on the SF might depend on the type and power of the jet \citep{Kalfountzou_17} and can even co-exist inside a galaxy \citep{Shin_19, Mercedes_23}. Feedback from stellar winds generated by massive SF has a similar ability to blow out the gas from a galaxy, even in the local Universe \citep{Chen_10}.

Mergers can trigger the AGN as they are able to funnel gas towards the galactic centre which also leads to a central starburst \citep{Hernquist_95, Sanders_96, Veilleux_02, Hopkins_06, Engel_10}. The SF boost is likely to be short-lived ($\sim$100 Myr) due to the depletion of the molecular gas and AGN feedback \citep[e.g.,][]{dimatteo_05}. Minor mergers can also have an impact on star-forming spiral galaxies, leading to SF quenching and a change in the morphology \citep[e.g.,][]{Bekki_98, Aguerri_01}. Conversely, in some cases, gas-rich minor mergers, have been shown to reignite recent SF in otherwise quiescent early-type galaxies \citep{Kaviraj_09}.

As mentioned above, intense SF in the galactic centre leads to a build-up of a large central spheroidal component. The bulge not only alters the morphology of a galaxy, but can also affect further star formation by stabilising the gas disk against fragmentation, lowering the efficiency of the SF process \citep[e.g.,][]{Martig_09, Ceverino_10, Bundy_10, Fang_13}. 

Galactic bars also appear to have a dual effect on SF. Gas in barred galaxies can lose angular momentum via gravitational torques, flow inwards \citep[e.g.,][]{Shlosman_90,Kim_11, Shin_17}. An increased central molecular gas concentration in barred galaxies has also been confirmed observationally \citep[e.g.,][]{Sakamoto_99, Sheth_05}. Some bars lead to the formation of a circumnuclear gas ring \citep[e.g.,][]{Colina_97}. Given the abundance of gas, a higher SFR is also expected in the centre of barred galaxies \citep[][]{Alonso_01, Hunt_08, Ellison_11, wang_12, catalan_17, Chown_19,Zee_23}; however, in some cases, the opposite is seen \citep{Sheth_05, wang_12, Wang_20}. This discrepancy is likely a consequence of the evolution of the bar. An increased SF activity could be attributed to the early stages of the bar growth, when there is much cold gas present \citep{Athanassoula_13}, but ultimately becomes depleted  \citep{Newnham_20} or rendered too turbulent to collapse \citep{Khoperskov_18}, which leads to quenching. Moreover, bar-driven gas inflows can occur multiple times during the evolution of a galaxy \citep{Jogee_05}.

Spiral structure in a galaxy is known for its ability to increase gas density, which favours SF \citep[e.g.,][]{Elmegreen_83, Elmegreen_94, kim_20}. There has been an ongoing debate whether the spiral structure is also able to boost the star formation efficiency (SFE) \citep{vogel_88, lord_90, cepa_90, knapen_96, seigar_02, gao_21} or not \citep{foyle_10, kreckel_16, schinnerer_17, Querejeta_21}. In addition, more tightly wound spiral arms are observed in redder, earlier type galaxies with higher stellar mass, and higher degree of central concentration \citep{yu_20}, and the spiral arms strength shows a positive correlation with SFR \citep{yu_21}.


The complex interplay between the processes described above is recorded by the star formation histories (SFHs) of galaxies. A possible approach to gauge SFHs is via spectral energy distribution (SED) fitting. This method is, however, proven to be rather complicated as real SFHs are very diverse, requiring a great deal of high-quality data and various model assumptions \citep[e.g.,][]{Papovich_01, Shapley_01, Muzzin_09, Conroy_13, Ciesla_16,Ciesla_17, Carnall_19, leja_19}, with short-term ($\sim$100~Myr) variations being especially difficult to pinpoint \citep[e.g.,][]{ocvirk_06, Gallazzi_09, Zibetti_09, leja_19}. A different approach that allows to study recent SHFs is done by comparing SF from the recent (5--10~Myr) and more distant (0.1--1~Gyr) past through observations \citep{Sullivan_00, Wuyts_11, Weisz_12, guo_16, Emami_19, Faisst_19, wolf_19, wang_lilly_20, Byun_21, Karachentsev_21} and simulations \citep{sparre_17, Broussard_19, flores_velazques_21} of galaxies. 

A common practise in this case is to probe recent SFHs by comparing H$\alpha$ line and  far ultraviolet (FUV) continuum emission. Since the bulk of H$\alpha$ emission originates from the recombination of hydrogen atoms ionised by young massive stars, it traces SF on timescales of $\lesssim$5-10~Myr (the typical main sequence lifetime of the O and B stars capable of ionising the gas). FUV observations, on the other hand, probe longer timescales of $\sim$100--200~Myr as the FUV radiation is emitted directly by young massive stars that have escaped their birth clouds \citep{kennicutt_evans_12}.

In \cite{lomaeva_22} (hereafter \citetalias{lomaeva_22}) we studied recent SFHs by calibrating a prescription that estimates the rate of change in the SFR at the present time, dSFR/dt. We aimed to use simple observables instead of full SED modeling of SFHs to avoid the strict data requirements and large computational costs associated with that.  Instead, we opted for a calibrated method that relies on as few observables as possible. This metric, \dsfr, which is the ratio between the SFR averaged over the past 5 and 200~Myr as a function of the H$\alpha-$FUV colour, was calculated via a set of models generated in \textsc{CIGALE} \citep[Code Investigating GALaxy Emission;][]{cigale_1, cigale_2, cigale_3}. 

In \citetalias{lomaeva_22}, we applied \dsfr\, to a grand-design nearby spiral NGC~628  seen almost perfectly face-on  \citep[inclination of 8.9$\degr$;][]{leroy_21_survey}. In this work, we extend our sample to 11 galaxies, including NGC~628, based on the observations within the PHANGS (the Physics at High Angular resolution in Nearby Galaxies project)\footnote{\url{https://sites.google.com/view/phangs/home/data}} ALMA \citep{leroy_21_pipeline, leroy_21_survey} and MUSE \citep{Emsellem_22} samples. The selected disk galaxies are star-forming, nearby (<20~Mpc), mostly face-on, with their optical images shown in the 1$^\textnormal{st}$ column of Figure~\ref{fig:all_maps}. They show a diversity of external (field vs. group/cluster) and internal (AGN vs. inactive; barred, unbarred, and disks without spirals) properties, which is perfect for examining recent changes in SFHs and potential physical processes behind them.

This paper is structured as follows: in Section \ref{sect:observations} we describe the observational data required for the analysis. In Section \ref{sect:method}, we briefly mention how the \dsfr\, metric was defined in \citetalias{lomaeva_22}. In Section \ref{sect:results}, we present the results showing \dsfr\, differences between the eleven galaxies sampled, the SFE behaviour within different internal galactic environments, and the difference in the distribution of \dsfr\, in the interarms vs. arms of some of the galaxies. Finally, in Section \ref{sect:discussion}, we discuss potential physical and artificial causes of the \dsfr\, differences in the arms-interarms as well as conclude the main findings in Section~\ref{sect:conclusions}.

\begin{figure*}
\begin{subfigure}{\textwidth}
	\includegraphics[width=\textwidth]{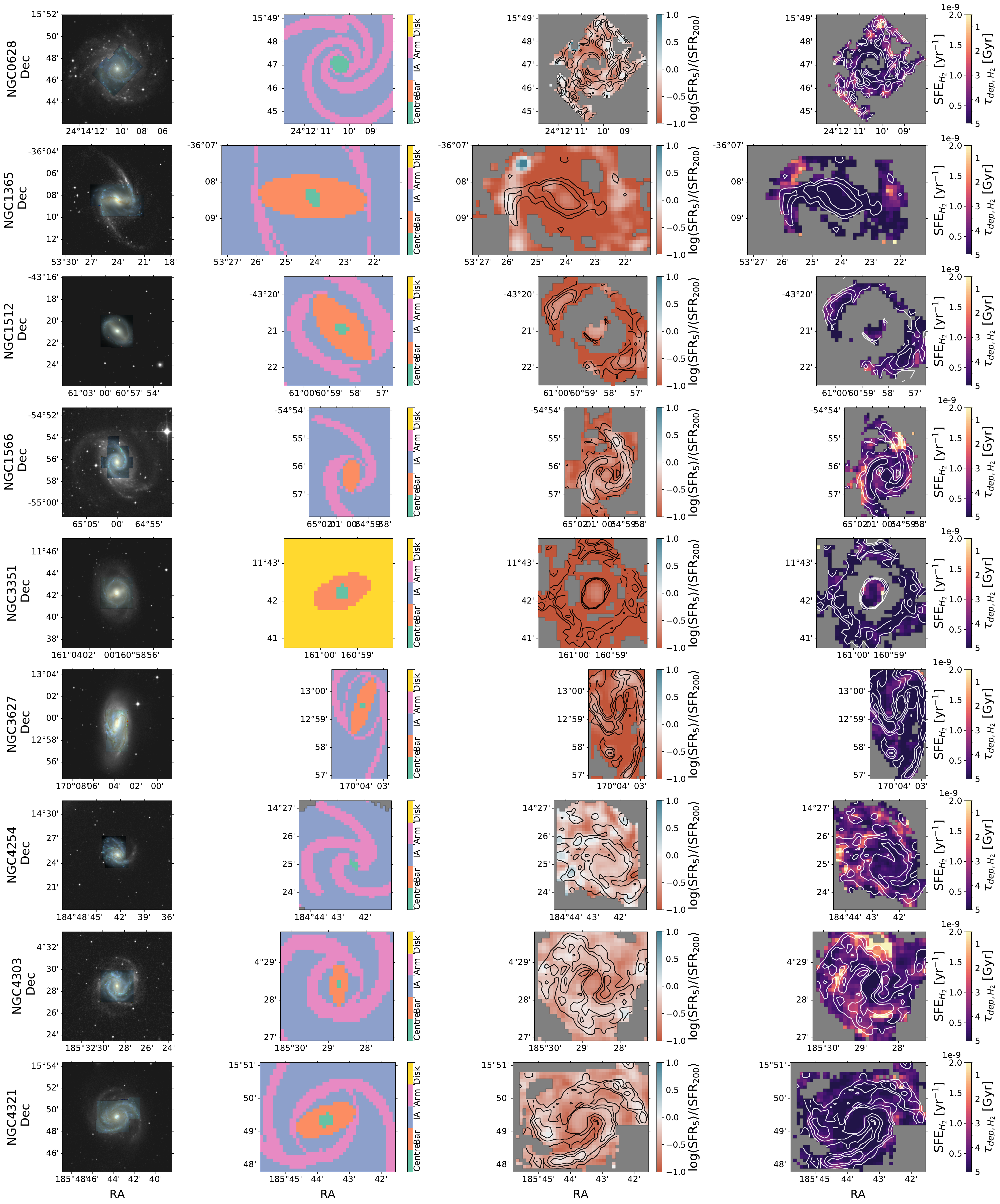}
\textcolor{white}{...}
\end{subfigure}
\end{figure*}

\begin{figure*}
\ContinuedFloat
\begin{subfigure}{\textwidth}
	\includegraphics[width=\textwidth]{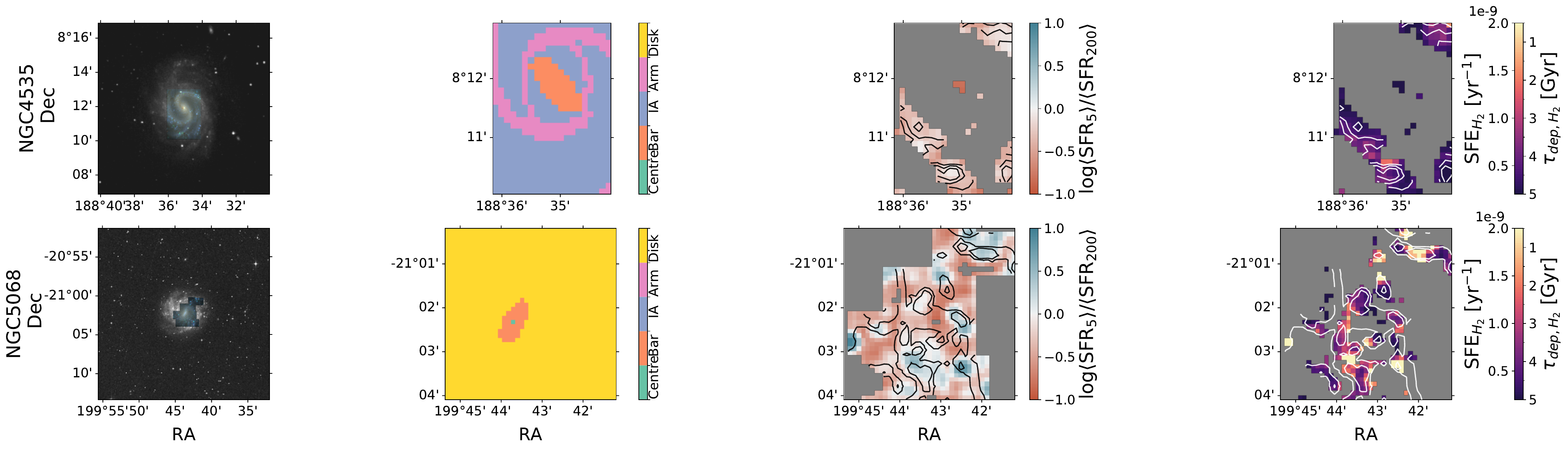}
\end{subfigure}
\caption{The eleven galaxies observed.\textit{ First column from the left}: an \textit{R}-band DSS image overlaid with a \textit{gri}-band photometric image from the PHANGS-MUSE sample \citep{Emsellem_22} to illustrate the region considered in this work. \textit{Second column}: the environmental mask from \protect \cite{Querejeta_21} convolved and rebinned to the 6\arcsec\, resolution and pixel size showing the central region, galactic bar, interarms, spiral arms, and disk without spirals if applicable. \textit{Third column}: the log\dsfr\, metric with the contours showing the $\Sigma_\textnormal{mol}$ concentration above the typical noise level.  \textit{Fourth column}: molecular SFE with the same contours as in the third column. The left-hand side of the colour bar shows SFE in the units of yr$^{-1}$, while the right-hand side exhibits the corresponding depletion time, $\tau_\textnormal{mol}$ in Gyr.}
\label{fig:all_maps}
\end{figure*}

\section{Observations}\label{sect:observations}

\subsection{Photometry}\label{sect:photometry}

Our analysis required photometric observations in three bands: far-UV (FUV), near- and mid-infrared (NIR and MIR, respectively). The FUV observations are required as an input for the \dsfr\,diagnostic, while the NIR and MIR data are used for correcting the FUV images for dust attenuation as well as for masking pixels with no recent SF (see Section~\ref{sect:optical_obs}). For those, we selected FUV images obtained by the GALaxy Evolution eXplorer \citep[GALEX;][]{galex, morrissey_07}, J-band images from the 2 Micron All-Sky Survey \citep[2MASS;][]{2mass}, as well as 3.6\micron\ and 24\micron\ images collected with the InfraRed Array Camera \citep[IRAC;][]{irac} and the Multiband Imager \citep[MIPS;][]{mips} on-board \textit{Spitzer}  \citep{spitzer}. The final sample selection in this work was dependent on the availability of the photometric data in these three bands among the PHANGS-MUSE galaxies, which left us with 11 objects.  

We downloaded the images from the DustPedia Archive\footnote{\url{ http://dustpedia.astro.noa.gr}} as well as the NED archive\footnote{\url{ http://ned.ipac.caltech.edu}}. The DustPedia sample contains matched aperture photometric images of 875 nearby galaxies in over 40 bands.

We also performed a check to estimate the sky background and removed it whenever applicable. We removed the foreground stars following the procedure in \citetalias{lomaeva_22}. Briefly, we identified the stars using the SIMBAD database \citep{simbad} and fitted a 2D Moffat profile to each star with the \textsc{mpdaf} package \citep{mpdaf1,Piqueras_19} in \textsc{Python}. With the FWHM of the Moffat profile, we centred a circular annulus and aperture at each star with the \textsc{photutils} \citep{photutils} package in \textsc{Python} and replaced the pixels inside the aperture with random numbers generated knowing the mean and one standard deviation of the immediately adjacent and sigma-clipped pixels inside the annulus. The sigma clipping was done  using \textsc{astropy} \citep{astropy1, astropy2} with a 2$\sigma$ threshold. The MIPS~24\micron\, data has the lowest resolution of 6\arcsec, hence, all other data products were convolved to this resolution. That is a slight improvement in resolution compared to \citetalias{lomaeva_22} where all observations were convolved to 7\arcsec.

The photometric observations were corrected for the Galactic extinction using the \cite{cardelli_89} dust extinction law with $E(B-V)$ values from \cite{Schlafly_11}, assuming the Galactic total-to-selective extinction value of $R_\text{V}$ = 3.1. 

The intrinsic UV luminosity emitted by the source, $L(UV)_\textnormal{int}$, is partly absorbed and re-emitted by the dust at infrared wavelengths. A common method to account for this effect is formulated by:
\begin{equation}
    L(UV)_\textnormal{int} = L(UV)_\textnormal{obs} + k \times L(IR)\,,
\end{equation}
where $L(UV)_\textnormal{obs}$ is the observed luminosity, $k$ is the scaling coefficient, and $L(IR)$ is the observed IR luminosity in the corresponding band.
\cite{leroy_19} derived a value for $k = 4.9$ using GALEX and Wide-field Infrared Survey Explorer (WISE) observations at 22~\micron\,and SED fitting from \cite{salim_16, salim_18}, which we employ in this work. This value was calibrated for a resolved sample of $\sim$15 750 nearby galaxies.

However, in \citetalias{lomaeva_22}, we corrected the FUV images for dust extinction following the recipe from \cite{boquien_16} instead. Briefly, \cite{boquien_16} performed a spatially resolved, multi-wavelength study of eight star-forming spiral galaxies from the KINGFISH survey \citep{kingfish}. Through SED fitting, the authors derived an equation connecting the $k$ coefficient and several photometric colours to estimate the variable impact of dust heated by old stellar populations (see their Table 4). The dust attenuation curves retrieved from numerical SED modelling tend to be steeper than the starburst curve inferred from empirical methods \citep[e.g.,][]{Calzetti_94, calzetti_00}. In our case, a curve that is significantly steeper than the star-burst \cite{calzetti_00} curve that we use to correct the H$\alpha$ emission for dust attenuation in Section~\ref{sect:optical_obs}, could add a systematic discrepancy. The recommendation of \cite{leroy_19} has also been used in other recent studies \citep[e.g.,][]{belfiore_23}, which makes it easier to compare our results with the recent literature. We further discuss these effects on our results in Section~\ref{sect:a_fuv_discussion}. For consistency, we explore the impact of both FUV attenuation methods using MIPS~24\micron\, observations. For $k$ derived as in \cite{boquien_16} we used:
\begin{equation}
    k = 15.044 - 2.169 \times \textnormal{(FUV--IRAC 3.6)}\, ,
\end{equation}
setting $L(IR) = L(\textnormal{MIPS 24}\micron)$ as listed as a possible IR band option in \cite{boquien_16}.

\subsection{Optical observations}\label{sect:optical_obs}

We used integrated line emission maps of the H$\alpha$, H$\beta$, [\ion{O}{iii}]$\lambda$5006~\AA, and the [\ion{Si}{ii}]$\lambda\lambda$6716, 6731~\AA\, doublet emission obtained with the MUSE spectrograph \citep[the Multi Unit Spectroscopic Explorer;][]{muse} at the Very Large Telescope (VLT). These observations were part of the PHANGS-MUSE survey \citep{Emsellem_22}. In total, the PHANGS-MUSE survey contains maps of 19 massive (9.4 < $\log$(M$_\star$/M$_\odot$) < 11.0) nearby (D $\lesssim$ 20~Mpc) star-forming disc galaxies. There are 168 MUSE pointings (1\arcmin\, by 1\arcmin\, each) with typical seeing ranges from 0.7-1\arcsec. Rather than using the data products at the native MUSE resolution, we opted for the data convolved and optimised (\textit{copt}), that is their point spread function (PSF) was  set to be a circular two-dimensional Gaussian with a FWHM that is constant as a function of wavelength and position within each mosaic. Those data products were subsequently convolved to 6\arcsec resolution to match the photometry. We applied a signal-to-noise (S/N) ratio cut of 5 to all optical observations. 

\cite{Emsellem_22} subtracted the sky background and corrected the MUSE data for foreground Galactic extinction using the \cite{odonnell_94} extinction curve and the $E(B-V)$ values from \cite{Schlafly_11}. Since the \cite{odonnell_94} extinction curve is only defined down to near-UV (NUV) wavelengths, we used the \cite{cardelli_89} curve instead that is available in the FUV for consistency. We also used the Galactic total-to-selective extinction value of $R_\text{V}$ = 3.1. 
The foreground stars were removed following the procedure described in Section~\ref{sect:photometry}.

Working with timescales as short as 5~Myr, that is with environments where dust and gas are mixed together, we needed to account for dust attenuation of the H$\alpha$ emission, $A(H\alpha)$. We calculated it for each spaxel from the Balmer decrement, F(H$\alpha$)/F(H$\beta$) assuming Case B recombination \citep{osterbrock_06}:
\begin{equation}
 \begin{split}
    A(H\alpha) \, \textnormal{[mag]} = \frac{E(H\beta - H\alpha)}{k(H\beta) - k(H\alpha)} \cdot k(H\alpha) = \\ = \frac{2.5\log \left( \frac{1}{2.86} \cdot \frac{F(H\alpha)}{F(H\beta)} \right)}{\frac{k(H\beta)}{k(H\alpha)}-1} \, ,
 \end{split}
\end{equation}
where $\frac{k(H\beta)}{k(H\alpha)}$  is the reddening curve ratio of 1.53 for the \cite{calzetti_00} curve assuming $R_\text{V}$ = 3.1 \citep{cardelli_89}. 

From this, we obtained the attenuation-corrected luminosity:
\begin{equation}
    L(H\alpha)_{\textnormal{corr}} = L(H\alpha)_{\textnormal{obs}} \times 10^{0.4 A(H\alpha)} \, ,
\end{equation}
where  $L(H\alpha)_{\textnormal{obs}}$ is the observed, attenuated H$\alpha$ luminosity.

Finally, we were able to calculate the SFR from the H$\alpha$ line emission assuming a \cite{kroupa_01} initial mass function (IMF), as
described in \cite{calzetti_summer}:
\begin{equation} \label{eq:sfr}
   \textnormal{SFR [M$_\odot$ yr$^{-1}$]}  = 5.5 \times 10^{-42} \times L(H\alpha)_{\textnormal{corr}} \, , 
\end{equation}
where $L(H\alpha)_{\textnormal{corr}}$ is in units of erg~s$^{-1}$.

\subsection{PHANGS-ALMA}

In addition to the optical observations, we also obtained $^{12}$CO(J=2$\rightarrow$1) line emission, hereafter CO(2–1), from the PHANGS-ALMA survey \citep[PI: E. Schinnerer;][]{leroy_21_pipeline, leroy_21_survey}. Limited by the MIPS~24\micron\, resolution, we used the line-integrated CO(2-1) intensity observations (broad mom0 map) at 7.5\arcsec\, resolution and a 1$\sigma$ sensitivity of 5.5~mJy~beam$^{-1}$ per 2.54~km~s$^{-1}$ channel. We applied an S/N ratio cut of 3 to the ALMA data.

To calculate the molecular gas mass surface density, $\Sigma_\textnormal{mol}$, from the line-integrated CO(2-1) intensity,  $I_{\textnormal{CO}(2-1)}$, in K~km~s$^{-1}$, we used the relation:
\begin{equation}
   \Sigma_\textnormal{mol}\, \textnormal{[M$_\odot$ pc$^{-2}$]} = \alpha^{1-0}_{\textnormal{CO}} \cdot R^{-1}_{21} \cdot I_{\textnormal{CO}(2-1)} \cdot \cos i \, ,
\end{equation}
where $\alpha^{1-0}_{\textnormal{CO}}$ is the CO(1-0) conversion factor in M$_\odot$ pc$^{-2}$ (K km s$^{-1})^{-1}$, $R_{21}$ is the CO(2-1)-to-CO(1-0) line ratio, and $i$ is the inclination. We assumed $R_{21}$=0.65, which is the non-weighted mean derived in \cite{den_brok_21}, and adopted a constant Galactic value for $\alpha^{1-0}_{\textnormal{CO}}$ of 4.35 M$_\odot$ pc$^{-2}$ (K km s$^{-1})^{-1}$, as in \cite{bolatto_13}.

\begin{table*}
\centering
\caption{The eleven galaxies studied in this work and some of their properties.
} 
\label{tab:sample_prop}
\begin{threeparttable} 
\begin{tabular}{lccccccccc}
\hline  
Name   & Type  & RA         & Dec     & $D$\tnote{a}  & $i$\tnote{b}   & $\log M_\star$\tnote{c} & $\log SFR$\tnote{c} & 12+log(O/H)\tnote{d} & Pixel scale\tnote{e} \\
       &       &[hms]       & [dms]   & [Mpc]         & [deg] & [M$_\odot$]    & [M$_\odot$ yr$^{-1}$]  & [dex] & [pc pixel$^{-1}$]  \\
\hline 
NGC 628 & Sc & 01h36m41.75s & +15d47m01.1s & 9.8 & 8.9 & 10.34 & 0.24 & 8.48 & 285.1  \\
NGC 1365 & Sb & 03h33m36.46s & -36d08m26.4s & 19.6 & 55.4 & 10.99 & 1.23 & 8.48 & 570.1  \\
NGC 1512 & Sa & 04h03m54.28s & -43d20m55.9s & 18.8 & 42.5 & 10.71 & 0.11 & 8.57 & 546.9 \\
NGC 1566 & SABb & 04h20m00.40s & -54d56m16.6s & 17.7 & 29.5 & 10.78 & 0.66 & 8.58 & 514.9 \\
NGC 3351 & Sb & 10h43m57.73s & +11d42m13.3s & 10.0 & 45.1 & 10.36 & 0.12 & 8.59 & 290.9    \\
NGC 3627 & Sb & 11h20m15.03s & +12d59m28.6s & 11.3 & 57.3 & 10.83 & 0.58 & 8.54 & 328.7   \\
NGC 4254 & Sc & 12h18m49.63s & +14d24m59.4s & 13.1 & 34.4 & 10.42 & 0.49 & 8.56 & 381.1  \\
NGC 4303 & Sbc & 12h21m54.93s & +04d28m25.6s & 17.0 & 23.5 & 10.52 & 0.73 & 8.58 & 494.5 \\
NGC 4321 & SABb & 12h22m54.93s & +15d49m20.3s & 15.2 & 38.5 & 10.75 & 0.55 & 8.56 & 442.1 \\
NGC 4535 & Sc & 12h34m20.34s & +08d11m51.9s & 15.8 & 44.7 & 10.53 & 0.33 & 8.54 & 459.6   \\
NGC 5068 & Sc & 13h18m54.81s & -21d02m20.8s & 5.2 & 35.7 & 9.40 & -0.56 & 8.32 & 151.3  \\
\hline
\end{tabular}

\begin{tablenotes}
  \item[a] \cite{Anand_21}
  \item[b] \cite{lang_20}
  \item[c] \cite{leroy_21_pipeline}
  \item[d] At r$_\textnormal{eff}$ from \cite{Groves_23}
  \item[e] Rebinned to 6\arcsec per pixel
  \end{tablenotes}
\end{threeparttable}
\end{table*}

\section{Method}\label{sect:method}
\subsection{\dsfr\,diagnostic}

In \citetalias{lomaeva_22}, we defined and calibrated an SFR change diagnostic, \dsfr, that allows to probe recent SFHs on timescales of 5--200~Myr and measure whether the SF has been increasing or decreasing in a given region inside a galaxy. The main advantage of this method is that it is based on simple observables instead of a full SED modeling. We tested this method on a well-known galaxy NGC~628, which is also included into the current study, with the purpose of applying this technique to a larger sample of star-forming galaxies in this work. 

In summary, the diagnostic is based on the H$\alpha-$FUV colour which is used as an input to calculate the ratio between the SFR averaged over the past 5 and 200~Myr. We used synthetic models generated with \textsc{CIGALE} \citep[Code Investigating GALaxy Emission;][]{cigale_1, cigale_2, cigale_3} to calibrate the metric and gauge potential degeneracies between H$\alpha-$FUV colour and different SFHs \citepalias[see][]{lomaeva_22}.

We calculated the H$\alpha$--FUV colour as:
\begin{equation}
     \textnormal{H$\alpha$--FUV} = -2.5 \cdot \log(\textnormal{LyC [mJy]}) + 20 - \textnormal{FUV [mag]}\, ,
\end{equation}
where $\log LyC$ is the Lyman continuum photons flux density that ionise \ion{H}{ii} regions derived from the H$\alpha$ luminosity, as presented and used in \cite{boselli_16, boselli_18, boselli_21}. For our sample, the median uncertainty in the H$\alpha$--FUV colour remains within 10\% for all galaxies.

In the next step, we generated a set of models in \textsc{CIGALE} that would represent hypothetical galaxies with varying levels of their recent SF. Briefly, we assumed a \cite{chabrier_03} IMF and stellar population synthesis models for the stellar emission from \cite{bruzual_03}. For the parametric SFH, we chose a delayed SFH with a constant instantaneous increase or drop in the SFR (\textsc{sfhdelayedbq}). Using the results from \cite{decleir_19}, who carried out SED fitting to individual pixels in NGC~628, we set the range of SFH parameters, such as the age, $A_\textnormal{main}$, and the e-folding time, $\tau_\textnormal{main}$, of the main stellar populations. The strength of the SFR increase/decrease was parameterised by $r_\textnormal{SFR}$, that is the ratio between the SFR after and before this event. As a result, $r_\textnormal{SFR} < 1$ represents a suppression in recent SF, $r_\textnormal{SFR} > 1$ denotes an increase in SF, and $r_\textnormal{SFR} = 1$ gives no change in the SFR. The lookback time in Myr of this SF boost/drop was set by $A_\textnormal{bq}$. Finally, we assumed a dust-free scenario in \textsc{CIGALE} and instead corrected the observed H$\alpha-$FUV colour for attenuation effects. All parameter ranges and a more detailed discussion on this procedure are presented in \citetalias{lomaeva_22} (Section 3). 

From the \textsc{CIGALE} models, we obtained a trend for \dsfr\, as a function of the H$\alpha-$FUV colour and fitted a curve to this trend using the \textsc{lmfit} package in \textsc{Python}. This yielded: 
\begin{equation}
    \begin{split}\label{eq:fit}
    \langle  SFR_{\textnormal{5}} \rangle \big/  \langle SFR_{\textnormal{200}} \rangle_\textnormal{fit} = \exp [ -10.123 \cdot \arctan (0.456 \cdot \\ \cdot (\textnormal{H}\alpha - \textnormal{FUV}))\, + 0.244 \cdot (\textnormal{H}\alpha - \textnormal{FUV}) + 7.820 ]\, .
    \end{split}
\end{equation}
 
In \citetalias{lomaeva_22}, we discussed two cases where the \textsc{CIGALE} models deviated from the main trend described by Equation~\ref{eq:fit}. In the first case, we saw a set of models where SF was quenched very strongly and abruptly, with $r_\textnormal{SFR} < 0.001$. We deem such scenarios unlikely in real galaxies. In the second case, recent changes in the SF activity could not be efficiently probed by the \dsfr\, parameter when no recent SF formation has occurred. To eliminate this quiescent scenario, we applied two additional constraints. The first one eliminates pixels that are too red in the FUV--IRAC~3.6\micron\, colour (> 3~mag), while the second one removes the regions where H$\alpha$ emission does not originate from SF, that is, the equivalent width of the H$\alpha$ line, EW(H$\alpha$), is too small (< 5~\AA). Since EW(H$\alpha$) observations were not available among the PHANGS-MUSE data products, we selected a different criterion, the NUV--J colour, such that  NUV--J~>~3.75~mag were masked out due to no recent SF. 

\section{Recent changes in the star formation activity}\label{sect:results}

\begin{figure}
	\includegraphics[width=\columnwidth]{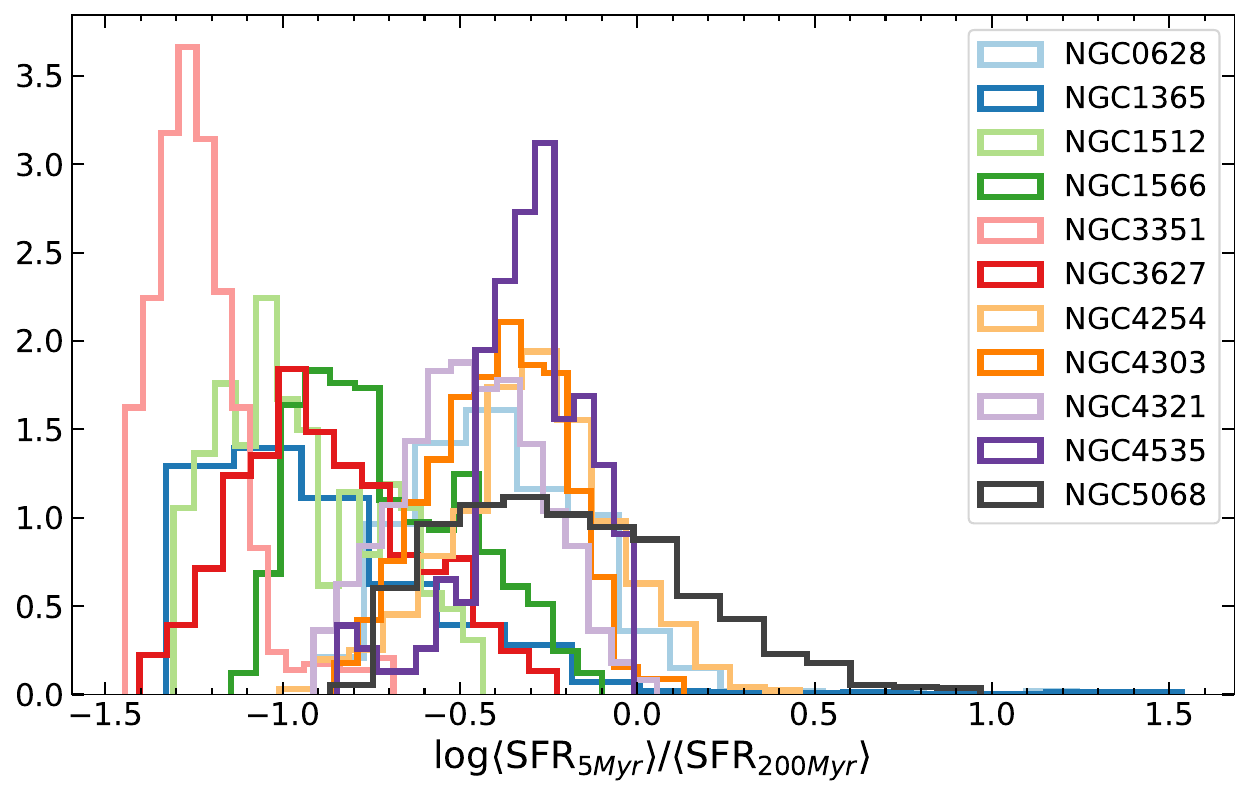}
    \caption{The distribution of log\dsfr\,in the observed galaxies, as defined in Equation \ref{eq:fit}. This shows that galaxies create two groups peaking at lower and higher values of \dsfr, as summariased in the last column in Table~\ref{tab:yes_no}. The histograms were normalised to have the same area of unity.}
    \label{fig:dsfr_distrib_by_gal}
\end{figure}

The distributions of \dsfr\, in each galaxy shown in Figure \ref{fig:dsfr_distrib_by_gal} tend to create two groups of galaxies peaking at higher (H) and lower (L) values of \dsfr. NGC~628, 4254, 4303, 4321, 4535, and 5068 belong to the high end of the \dsfr\, distribution, while NGC~1365, 1512, 1566, 3351, and 3627 belong to its low end.This is also summarised in the last column of Table~\ref{tab:yes_no}.

Looking at the \dsfr\, distribution across the galactic disks in the 3$^{\textnormal{rd}}$ column of Figure \ref{fig:all_maps}, we observe large areas of increasing recent SF in NGC~628, 4254, 4303, and 4321, while NGC~1512, 1566, and 3627 exhibit recent SF activity that ranges from roughly unchanged to decreasing. The most active galaxy in terms of recent SF changes is the low-mass galaxy NGC~5068. NGC~1365 shows a region with a very strong recent SF increase, visually similar to the headlight cloud in NGC~628 \citep{headlight_cloud}. NGC~3351 is the only galaxy in the sample that has a consistently decreasing recent SF across the entire disk observed. Finally, NGC~4535 shows a slight increase in the recent SF in the arm regions, although a large fraction of the observed pixels is masked out due to the limitations of the H$\beta$ line SNR. 

\subsection{The impact of AGN, bars, and external effects on recent SF}\label{sect:dsfr_env_breakdown}

In this section, we examine if there is any connection between a galaxy's position in the \dsfr\, distributions, AGN, galactic bar, vicinity to other galaxies, and past gravitational interactions with other group/cluster members. We summarise these properties in Table~\ref{tab:yes_no}.

\begin{table*}
\centering
\caption{The summary of the internal and external properties of the eleven galaxies studied in this work. References for the columns AGN, Field, and Interacting can be found in Section \ref{sect:dsfr_env_breakdown}. The \dsfr\ column denotes if a galaxy belongs to the high (H) or low (L) end of the \dsfr\ distribution in Figure~\ref{fig:dsfr_distrib_by_gal}.} 
\label{tab:yes_no}
\begin{tabular}{lccccc}
\hline  
Name   &  AGN & Bar & Field & Interacting & \dsfr \\
       &  [Yes/No] & [Yes/No]  & [Yes/No] & [Yes/No] & [Low/High]  \\
\hline 
NGC 628  & N & N & N  & N  & H \\
NGC 1365 & Y & Y & N  & N  & L \\
NGC 1512 & N & Y & N  & Y  & L\\
NGC 1566 & Y & Y  & N  & Y  & L\\
NGC 3351 & N & Y  & Y  & N  & L  \\
NGC 3627 & Y  & Y & N  & Y  & L \\
NGC 4254 & N  & N & N  &  Y & H\\
NGC 4303 & Y & Y & N &  N  & H \\
NGC 4321 & Y & Y  & N & N  & H \\
NGC 4535 & N & Y  & N & N  & H  \\
NGC 5068 & N  & Y & Y  & N  & H \\
\hline
\end{tabular}
\end{table*}

\subsubsection{AGN feedback}

Among AGN hosts in the sample, there are Seyfert galaxies NGC~1365  \citep{veron_06} and NGC~1566 \citep{Shobbrook_66}, low ionization nuclear emission region (LINER)/Seyfert type 2 galaxies NGC~4303 and NGC~3627 \citep{agn_3627}, and the starburst NGC~4321 which is another LINER galaxy \citep{Wozniak_99}.

For NGC~1365, \cite{gao_21} showed that the feedback from the AGN or hot shocked gas can push away molecular and atomic hydrogen gas at a speed up to 100~km s$^{-1}$. They saw that the outflow of the molecular gas is faster than the SF, suggesting a negative feedback scenario. \dsfr\, indicates a drop in the SF activity in the central region of NGC~1365, in agreement with this statement.

In NGC~1566, the super-massive black hole influences the gas dynamics reverting the gravity torques, as shown in \cite{Combes_14}. This drives the gas inwards and possibly fuels the AGN.

Galaxies with no confirmed or debated AGN include NGC~628 \citep{Liu_05, alston_21}, NGC~1512 \citep{Ducci_14}, NGC~3351  \citep{Gadotti_19}, NGC~4254 \citep{Burtscher_21}, NGC~4535\footnote{\label{footnote1}As per NED \url{https://ned.ipac.caltech.edu/}}, and NGC~5068\footnote{See Footnote \ref{footnote1}}.

We also note that it is likely that the presence of an AGN would not have a substantial effect on the SF activity, unless the AGN duty cycle is much longer than typically assumed. This is a result of the large differences between the dynamical timescales of the AGN and the galaxy as a whole.

\subsubsection{Galactic bar presence}
There are nine galaxies in the sample that have a galactic bar, as per the environmental masks in \cite{Querejeta_21}: NGC~1365, 1512, 1566, 3351, 3627, 4303, 4321,  4535, and 5068, while NGC~628 and NGC~4254 have no bar. 

NGC~1365 is dubbed the Great Barred Spiral Galaxy and has the highest global SFR and stellar mass in our sample, with most of the SF occurring in the central starburst ring. \cite{Schinnerer_23} found that the large-scale stellar bar in NGC~1365 drives gas inwards and might be solely responsible for the massive SF in the central 5~kpc, without any significant AGN feedback on the gas disk.

NGC~1512 has a double ring structure with one ring located around the nucleus and another one situated in the disk, which is not uncommon in barred galaxies. \cite{ma_17, ma_18} estimated the age of the nuclear ring in NGC~1512 to be $\sim$40~Myr on average. The \dsfr\, metric indicates a moderate decline in the recent SF activity in the central part of the galaxy, in line with these findings. NGC~1566 also has a double ring structure, where the rings are formed by the wound-up spiral arms. NGC~3351 contains a circumnuclear ring as well with intense massive SF \citep[e.g.,][]{Alloin_82, Colina_97}  and an inner ring at a larger radius. Previous studies have revealed the gas and dust in the circumnuclear ring are accumulated via the bar \citep[e.g.,][]{Leaman_19}.

\cite{Iles_22} simulated interacting and isolated bars to match NGC~3627 and NGC~4303, respectively. They found that the presence of a bar boosts SF activity in both cases, however, occurring gravitational interactions cause an additional burst-like period of the SF, likely due to the consolidation of the spiral arm before the bar forms. Indeed, we see that \dsfr\, decreases more in NGC~3627 than in NGC~4303, which could be due to a stronger gas depletion in NGC~3627 during the earlier SF bursts.

\subsubsection{Cluster/group membership}

There are two galaxies in the sample that are isolated filed galaxies: NGC~3351 \citep{Young_96} and NGC~5068 \citep{Karachentsev_02}. 

Among the galaxies that belong either to a group or a cluster, there are NGC~628, 1365, 1512, 1566, 3627, 4303, 4254, 4321, and 4535. 

NGC~628 is a member of the NGC~628 group \citep{Garcia_93}. NGC~1365 is situated in the Fornax cluster \citep{Lindblad_99}, while NGC~4303, 4254, 4321, and 4535 are located in the Virgo Cluster \citep{Binggeli_85, Fouque_92}. NGC~1512 and NGC~1566 are members of the Dorado group \citep{Maia_89}. Finally, NGC~3627 is a galaxy in the Leo Triplet \citep{Ferrarese_00}.

We caution, however, that a more thorough examination of the group and cluster membership would require a consideration about its size or density as well as a galaxy's position in it.

\subsubsection{Past gravitational interactions}
    
Among interacting galaxies, there is NGC~1512, which is interacting with the dwarf galaxy NGC~1510. This galaxy pair is separated only by $\sim$5\arcmin\,(13.8 kpc) \citep{Koribalski_09}. NGC~1566 is also known to interact with its smaller companions \citep{Kilborn_05}. NGC~3627 has potentially interacted with the neighbouring NGC~3623 and NGC~3628 in the past \citep{Haynes_79}. NGC~4254 is likely to have experienced a rapid gravitational encounter with another cluster member 280–750~Myr ago \citep{Vollmer_05, Boselli_18b}. NGC~4321 has probably interacted with other cluster companions in the past as well \citep[e.g.,][]{Knapen_93}.

The non-interacting galaxies include NGC~628, 1365, 3351, 4303, 4321, 4535, and 5068. NGC~628 is considered isolated since it has not experienced any gravitational interactions in the past 1~Gyr \citep{Kamphuis_92}. This view has, however, been recently challenged \citep{Wezgowiec_22}. NGC~4535 is rather unperturbed, however, might be interacting with the cluster environment, potentially experiencing ram pressure stripping \citep{Wezgowiec_07}.

Overall, we see that the recent changes in the SF are difficult to pinpoint to the extragalactic environment, past and current gravitational interactions with other galaxies as well as the presence of an AGN and a galactic bar. We also bear in mind the low sample size in this study.

\subsection{Effects of the internal galaxy environment on SFE}\label{sect:results_sfe}

\begin{figure*}
	\includegraphics[width=\textwidth]{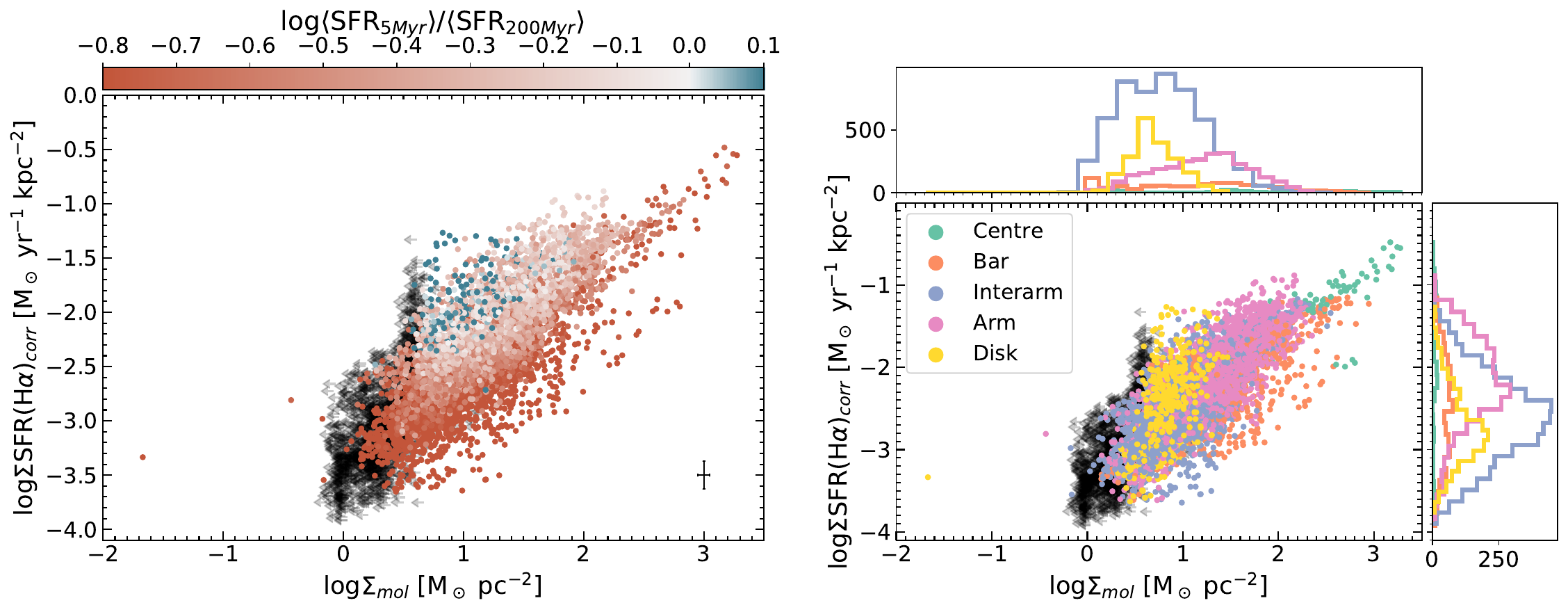}
    \caption{The resolved Kennicutt-Schmidt relation (rKS) of the galaxies sampled. The black arrows denote upper limits. \textit{Left panel}: the rKS coloured by the \dsfr\, metric, with blue and red data points indicating increased vs. decreased recent SF activity, respectively. The error bar represents the mean uncertainty among the galaxies for detected regions above the S/N cut. \textit{Right panel}: the rKS coloured according to the different internal galactic environments taken from \protect \cite{Querejeta_21}. The top and right-most histograms show the respective distributions of $\Sigma_\textnormal{mol}$ and $\Sigma_\textnormal{SFR(H$\alpha$)}$ in the internal environments considered.}
    \label{fig:rks}
\end{figure*}

Similarly as in \citetalias{lomaeva_22}, we constructed a resolved Kennicutt-Schmidt relation, as shown in Figure~\ref{fig:rks}. In the left panel of Figure~\ref{fig:rks}, we see a gradient in \dsfr\, for fixed $\Sigma_\textnormal{mol}$. This gradient is somewhat steeper compared to the one of NGC~628 in \citetalias{lomaeva_22} since we chose a different method to correct for the FUV attenuation. A number of pixels with $\log$\dsfr\, > 0 are positioned at an intermediate value of $\log\Sigma_\textnormal{mol}$ $\approx$ 0.9 M$_\odot$ pc$^{-2}$ that belong to the low stellar mass galaxy NGC~5068. Low-mass galaxies do tend to experience a ``burstier'' SF \citep[e.g.,][]{Weisz_12, Emami_19, flores_velazques_21}, which, however, does not explain the high \dsfr\, and SFE at the intermediate $\log\Sigma_\textnormal{mol}$. Instead, we might actually underestimate $\log\Sigma_\textnormal{mol}$ in NGC~5068: having lower gas phase metallicity, A(FUV), and A(H$\alpha$) than the other galaxies in the sample, the assumption of a constant Galactic $\alpha_\textnormal{CO}=4.35$~M$_\odot$ pc$^{-2}$~(K km s$^{-1})^{-1}$ is likely to be too low. It is known that  CO molecules can become UV photo-dissociated in galaxies with a lower metal and dust content, while the H$_2$ reservoir remains self-shielded \citep[e.g.,][]{amorin_16, madded_20}. A higher $\alpha_\textnormal{CO}$ would push the blue data points associated with NGC~5068 to higher $\log\Sigma_\textnormal{mol}$ values in Figure~\ref{fig:rks}, rendering them more aligned with the overall trend. For example, the gas phase metallicity of $Z=8.32$\footnote{ The oxygen abundance in units of 12+log(O/H)} in NGC~5068 yields $\alpha_\textnormal{CO}=17.7$, following the recipe from \cite{Accurso_17}. We do not opt for any variable $\alpha_\textnormal{CO}$ calibration here since it would introduce additional scatter to the analysis, while the correct approach to calculate $\alpha_\textnormal{CO}$ would also consider its radial variations, which is beyond the scope of this work.

One of the exercises of this work was to examine how different morphological structures inside a galaxy affect the SF. To separate each galaxy into the bulge, bar, arm and interarm regions, or alternatively disks without spirals, we applied a simple environmental mask from \cite{Querejeta_21}\footnote{\url{http://dx.doi.org/10.11570/21.0024}}. These masks were constructed from \textit{Spitzer} IRAC~3.6\micron\, images at $\sim$1.7\arcsec\, resolution and later convolved and rebinned to 6\arcsec\, resolution and pixel size to match the other observations. 

Looking at the right-hand panel in Figure~\ref{fig:rks}, we see that the interarm regions tend to have lower $\log\Sigma_\textnormal{mol}$ and $\log\Sigma_\textnormal{SFR(H$\alpha$)}$ compared to the arm regions. These results are in line with expectations that spiral arms tend to concentrate molecular gas which also increases the SFR \citep[e.g.,][]{kim_20} compared to the interarms. Galaxies, such as NGC~3351 and NGC~5068 that are classified as disk galaxies without spirals, show a similar SFR and molecular gas surface density distribution in the disk as the interarms regions in galaxies that do exhibit a spiral structure. The bars and galactic  centres have overall a wide distribution of $\log\Sigma_\textnormal{mol}$ and $\log\Sigma_\textnormal{SFR(H$\alpha$)}$, with the central pixels reaching out to the highest values of these quantities. These pixels are attributed to a single galaxy, NGC~1365. This galaxy is known for its intense SF inside the central ring-like structure which is formed by the gas inflows driven by the large-scale stellar bar \citep{Schinnerer_23}. Galactic bars and bulges can have varying effects on the SFR and SFE due to complex gas dynamics \citep[e.g.,][]{Querejeta_21, Iles_22, maeda_23}, hence the large spread of these pixels. 

\begin{figure*}
	\includegraphics[width=\textwidth]{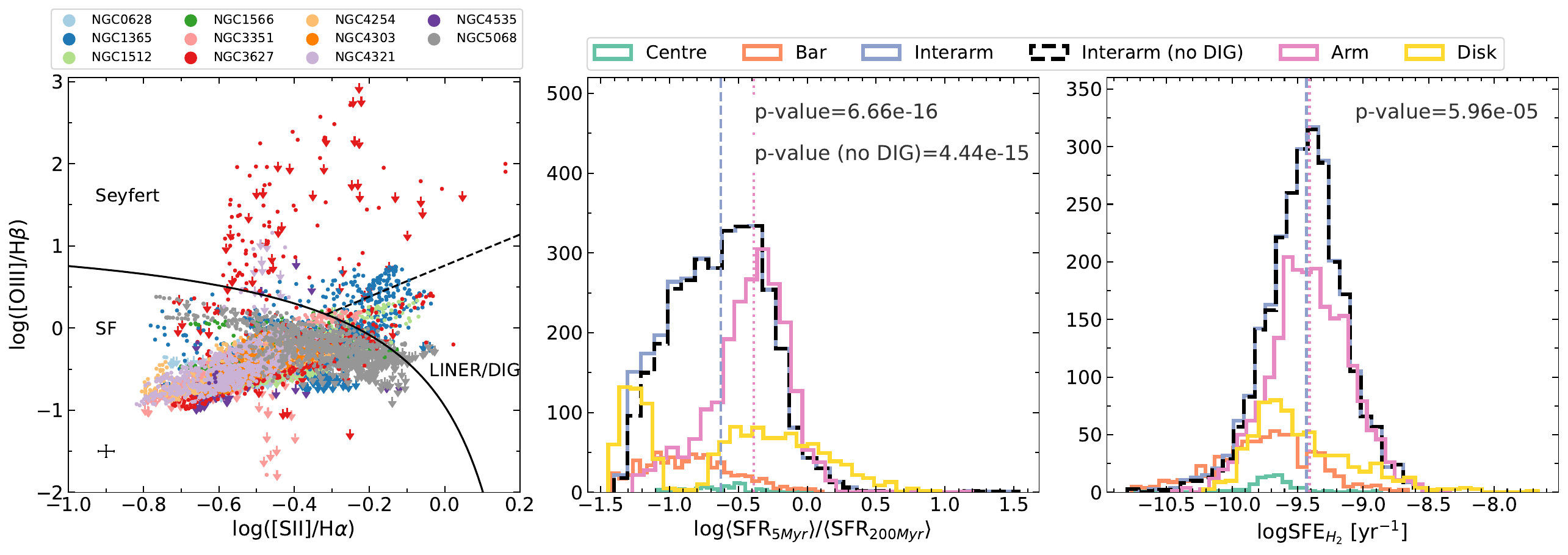}
    \caption{\textit{Left panel}: the BPT diagram of the galaxies sampled used to identify the DIG regions. The error bar represents the mean uncertainty among the galaxies for the detections above the S/N cut.\textit{Middle panel}: the log\dsfr\, distributions in the different internal galactic environments. The p-values are from the K-S test performed on the interarms vs. arms \dsfr\, distributions, both for the full samples and with the DIG associated pixels excluded. The lines represent median values of the arms and interarms. \textit{Right panel}: same as in the middle panel, but for the molecular SFE. }
    \label{fig:bpt}
\end{figure*}

We also illustrate molecular SFEs (SFE$_\textnormal{H$_2$} = \Sigma_\textnormal{SFR(H$\alpha$)}/ \Sigma_\textnormal{mol}$) in different internal environments in the rightmost panel of Figure~\ref{fig:bpt} by plotting their distributions of the molecular SFE. We are also interested if these trends are affected by H$\alpha$ emission unrelated to the SF, such as the emission from the diffuse ionised gas (DIG). The DIG contribution is especially important in the interarm region where the SF is generally low. To identify such pixels we constructed a Baldwin, Phillips, Terlevich \citep[BPT;][]{bpt} diagram with the extreme starburst classification line computed by \cite{Kewley_01} and the Seyfert and low-ionization narrow emission-line (LINER) separation from \cite{Kewley_06}. 

The arm-interarm regions peak at somewhat higher SFE values compared to the bulges, bars, and disks without spirals. The removal of the DIG pixels does not change the SFE distribution in the interarms, similarly as for \dsfr, as mentioned further in Section~\ref{sect:results_dsfr}. Nevertheless, the Kolmogorov-Smirnov test (K-S test) yielded a statistically significant p-value, meaning that the arm and interarm samples of the SFE are intrinsically different. This actually disagrees with the corresponding results for NGC~628 in \citetalias{lomaeva_22}, where we found that the SFE in the arms and interarms was similarly distributed.

There have been a number of studies examining whether the SFE in the galactic bars and disks are different. Some works have seen a suppression of the SFE in bars despite available molecular gas \citep{Momose_10, Maeda_20, maeda_23}, while other works have not observed any such SFE suppression \citep{Saintonge_12, diaz_21, Querejeta_21, Muraoka_19}. In our sample, we find that the SFE$_\textnormal{bar}$/SFE$_\textnormal{arm}$ = 0.44, while the K-S test yields a p-value consistent with zero, meaning that the SFE distribution in the arm and bar are intrinsically different. Thus, we observe a suppression of the SFE in the bar compared to the spiral arms. 

Generally, the SFE appears rather similar in the galactic bars, disks without spirals, and inside the central regions, while it is slightly higher on average in the arm-interarm regions, as shown in the right panel of Figure~\ref{fig:bpt}. The SFE distributions in the arms and interarms show a somewhat different shape and peak widths, with the the K-S test showing that they are statistically different. Their median values are, however, very similar.

\subsection{\dsfr\, distribution in the arms and interarms} \label{sect:results_dsfr}

Similarly as for the molecular SFE in Section~\ref{sect:results_sfe}, we investigated the distribution of the log\dsfr\, parameter in different internal galactic environments, as shown in the middle panel of Figure~\ref{fig:bpt}. This panel shows that the \dsfr\ distribution in the arms and interarms covers the high distribution end very similarly, while the low distribution end is populated by the areas associated with the interarms much more. This observation is not affected by the DIG regions, as shown in the same figure. 

We also performed the K-S test for \dsfr, and it produced a similar result as for the SFE, albeit with a higher significance. That is in line with the previous observations for NGC~628 in \citetalias{lomaeva_22}. In addition to having different shapes, we also see that the arm-interarm \dsfr\, distributions have drastically different medians.

\begin{figure}
	\includegraphics[width=\columnwidth]{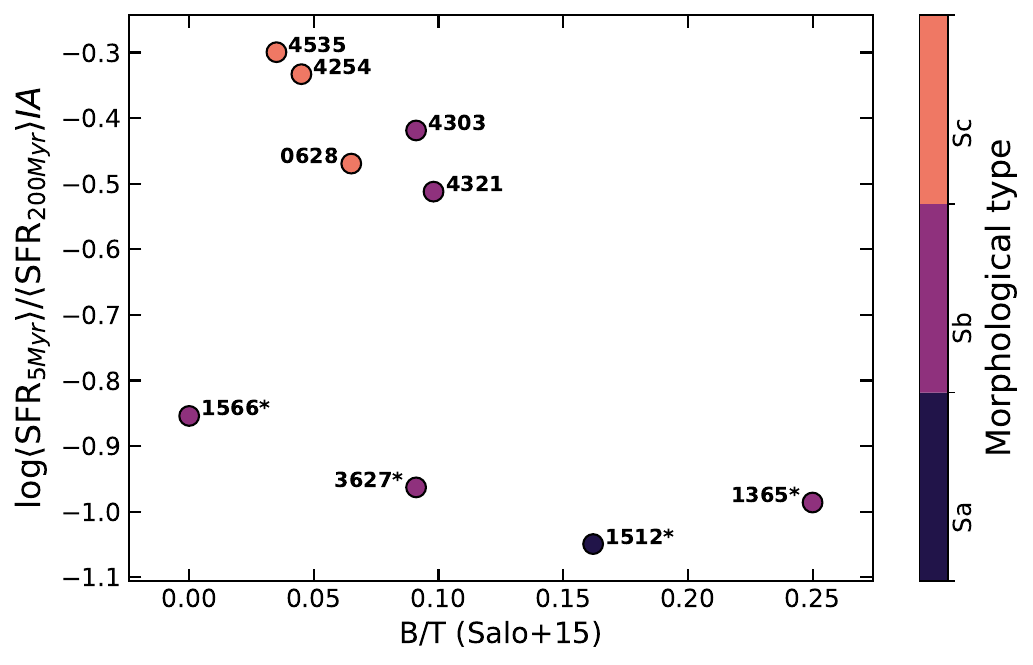}
    \caption{The median log \dsfr\ in the interarms as a function of B/T values from \protect\cite{salo_15}, colour-coded according to the morphological type, as in Table~\ref{tab:sample_prop}, however, without a distinction between the Sb and SABb types. The galaxies with lower median log \dsfr\ in the interarms, i.e., those exhibiting larger differences in \dsfr\ in the arms-interarms (marked with an asterisk), tend to be of an earlier type and have larger B/T values than the galaxies with more similar recent SFHs in the arms-interarms. }
    \label{fig:dsfr_morph_type}
\end{figure}

In an attempt to understand the differences in the \dsfr\, distribution in the arms and interarms, we looked into the global properties of the galaxies, such as the bulge-to-total light ratio (B/T) from \cite{salo_15} and morphological type, as shown in Figure~\ref{fig:dsfr_morph_type}. Those galaxies where the median \dsfr\ in the interarms is lower, resulting in larger differences in \dsfr\ in the arms-interarms (dubbed different A-IA as opposed to similar A-IA), tend to have a higher B/T ratio, as opposed to the similar A-IA galaxies. Moreover, we also see that many of the similar A-IA galaxies belong to a later morphological type (Sc), again shown in Figure~\ref{fig:dsfr_morph_type}, while the majority of the different A-IA  galaxies belong to the Sa and Sb type, which are characterised by a larger bulge and less prominent spiral structure compared to the Sc type. Therefore, our results also demonstrate that the galaxies with larger B/T ratios are of an earlier morphological type, which is supported by their larger bulges contributing a higher fraction of the total emitted light. We note however, that this sample size is too small to make an exhaustive conclusion about the trend.

In summary, we notice a characteristic behaviour among the galaxies in the sample where some have \dsfr\, stretching down to lower values in the interarms  and show a more distinct difference from the \dsfr\ distribution in the arms, while other galaxies have rather similar \dsfr\ in the same regions. The former group also tends to have galaxies with larger B/T values and an earlier morphological type.

\section{Discussion}\label{sect:discussion}

\subsection{Potential influences on \dsfr}
In this section, we discuss possible explanations to the observed differences between the \dsfr\, distributions in the arms and interarms. We try to understand why the interarm regions have undergone a somewhat more diverse recent changes in the SF compared to the spiral arms, as indicated by the \dsfr\, distribution in the middle panel of Figure~\ref{fig:bpt}.

\subsubsection{Morphology and global stellar mass}

As mentioned, the galaxies with lower median \dsfr\ that also have more distinct recent SF changes in the arms-interarms tend to have larger stellar masses and earlier morphological types, as shown in Figure \ref{fig:dsfr_morph_type}. Previous studies suggest that massive galaxies with larger bulges can have low star formation activity, as the gas can be affected by feedback or stabilised against collapse \citep[e.g.,][]{Martig_09, Peng_10, Fang_13}. The reach of these potential bulge-driven processes is, however, unclear and might have a limited effect on the interarm regions, given the scales of several kpc studied here. 

In addition, \cite{Elmegreen_11} found that the contrast between arms-interarms tends to increase with later Hubble type (from flocculent to multiple arm to grand design) using NIR photometry, while \cite{Savchenko_20} observed in optical/NIR a somewhat larger arm width in grand design spirals in their analysis of these galaxy types. Another recent work by \cite{Stuber_23} who studied PHANGS-ALMA galaxies found that their classification agrees better with the optical/NIR literature definitions for grand design (67\% of cases) than for multi-arm (47\% of cases) galaxies. Thus, the spiral structure in earlier type galaxies tends to be more wavelength dependent. Therefore, for early-type galaxies in our sample that have lower arm-interarm contrast, narrower spiral arms, and higher wavelength dependency we may be contaminating the true \dsfr\ signal in the arms-interarms as our morphological masks derived in the NIR are more likely to be imprecise.

\subsubsection{$\Sigma_\textnormal{mol}$ and SFE}

A potential explanation to the differences in \dsfr\, between the two A-IA populations could be attributed to the molecular gas availability. 
The top histogram in right panel of Figure~\ref{fig:rks} hints at $\Sigma_\textnormal{mol}$ being shifted towards lower values in the interarms than in the arms. However, when we separate the different and similar A-IA, we do not see any conclusive differences in the $\Sigma_\textnormal{mol}$ distributions.

Regarding SFE, the right-most panel of Figure~\ref{fig:bpt} illustrates that it has as similar range and shape in the arms and interarms, having almost identical median values. Thus, we cannot conclude if the differences in the recent SFHs in the arms and interarms are driven by the gas content availability or SFE. This points towards other factors preventing SF from taking place.

\subsection{Model caveats}

\subsubsection{A(FUV) corrections}\label{sect:a_fuv_discussion}

In addition, we investigated how the choice of a constant or a variable scaling coefficient $k$ used for correcting FUV emission for dust extinction affects the distribution of \dsfr\, in the arms and interarms. We used a constant $k=4.9$ from \cite{leroy_19} who derived the scaling coefficient using SED fitting for a sample of  $\sim$15 750 galaxies at $z=0$. The variable value was estimated with the \cite{boquien_16} recipe who also performed SED fitting in \textsc{CIGALE} for eight nearby galaxies. Both methods are applicable for resolved studies. The distribution of the variable vs. constant $k$ are shown in the left panel of Figure~\ref{fig:bpt_constant_k}. 

The constant value is considerably lower than the variable $k$, which ranges mostly between $k$~=~4--12. This is in agreement with the findings in \cite{belfiore_23} who obtained a result $\sim$0.3~dex lower for the \cite{leroy_19} vs. \cite{boquien_16} recipes. As mentioned before, this could be attributed to a steeper attenuation curve in \cite{boquien_16}. \cite{decleir_19} similarly obtained a rather steep attenuation curve from SED fitting in \textsc{CIGALE} for NGC~628 (see their Figure 8). 

\begin{figure*}
	\includegraphics[width=\textwidth]{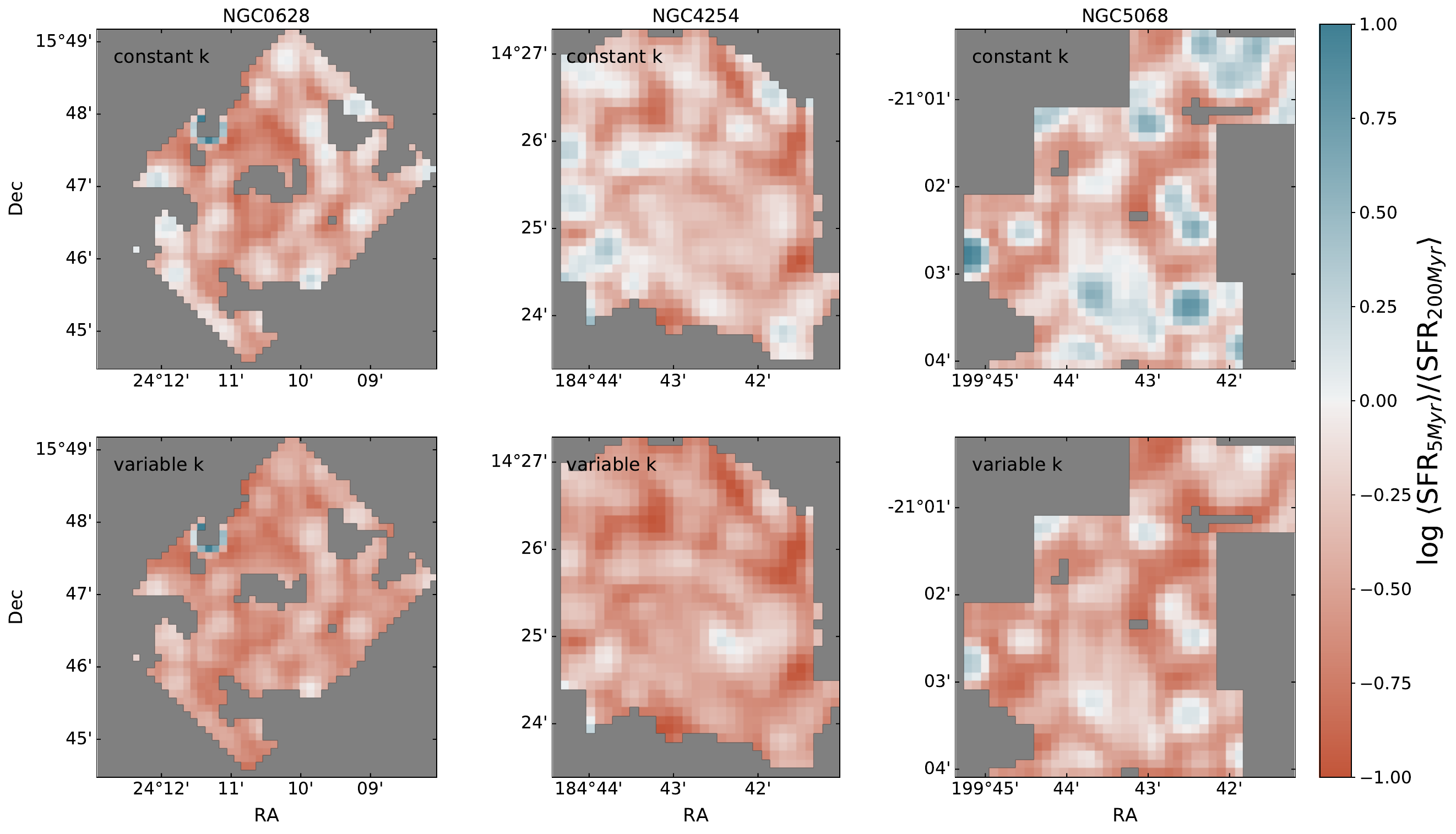}
    \caption{The effects on \dsfr\, when implementing a constant $k=4.9$ from \protect\cite{leroy_19} (top row) and a variable $k$ from \protect\cite{boquien_16} (bottom row) in A(FUV) calculations.}
    \label{fig:dsfr_const_k}
\end{figure*}

\begin{figure*} 
	\includegraphics[width=\textwidth]{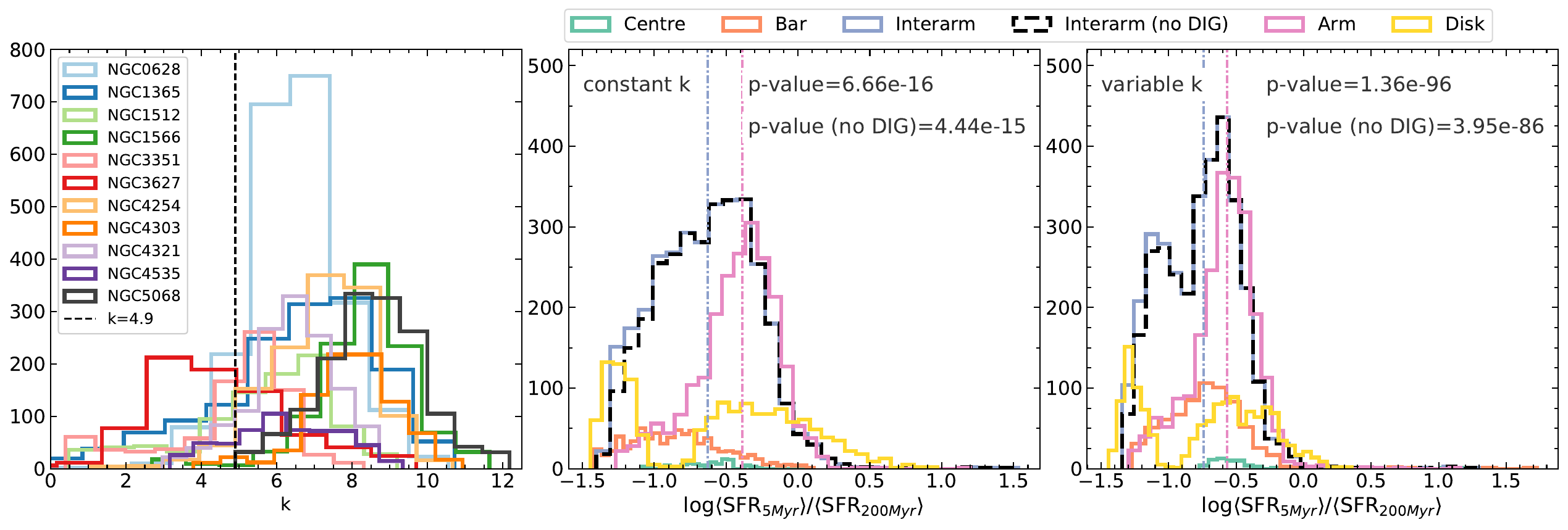}
    \caption{\textit{Left panel}: variable $k$ from \protect\cite{boquien_16} in colour vs. the constant $k$ (dashed line) from \protect\cite{leroy_19}. \textit{Middle panel}: the logarithm of \dsfr\, when A(FUV) is corrected assuming $k=4.9$ from \protect\cite{leroy_19}. The dot-dashed lines denote median values for the arm and interarm distributions. The p-values were derived using the K-S test. \textit{Right panel}: same as in the middle panel but with the variable $k$ from \protect\cite{boquien_16}.}
    \label{fig:bpt_constant_k}
\end{figure*}

Nevertheless, applying the constant value of $k=4.9$ has an effect on the \dsfr\, metric as it increases the observed dynamic range of \dsfr, as shown in Figure~\ref{fig:dsfr_const_k}. This is expected because $k$ is supposed to be enhanced in star-forming regions, as opposed to more quiescent areas where the dust is mostly heated by older stars. Thus, a constant $k$ is likely to be larger than its true value in quiescent regions and lower in star-forming ones. Since the variable $k$ values are mostly larger than the constant one, A(FUV), and subsequently L(FUV)$_\textnormal{corr}$, becomes lower in the latter case, increasing \dsfr. With a constant $k$, we observe a higher number of pixels that have experienced a recent increase in SF compared to the variable case and \citetalias{lomaeva_22}. That is logical, considering that on average the changes in the SFR measured on the short scales of $\lesssim$200~Myr over the galactic disc should be unchanged, with the total \dsfr\, of approximately zero. 

We are also interested whether the $k$ coefficient changes the distribution of \dsfr\, in the arms and interarms. The middle and right panels of Figure~\ref{fig:bpt_constant_k} show that the overall distribution of \dsfr\, in the interarms is preserved regardless of the choice of $k$. Thus, we deem the use of a constant value for $k$ a more suitable choice in our case since it reduces the systematic discrepancies between the shape of the attenuation curves used to correct for the FUV and H$\alpha$ attenuation. 

\subsubsection{SFR measurements in faint regions}

The analysis in this works relies on measurements of the SFR, and thus, assumptions made about the shape of the IMF. 
The interarm regions are characterised by a lower $\Sigma_\textnormal{SFR(H$\alpha$)}$ than the spiral arms, and since we are examining characteristic differences between the two environments, we need to ensure that these distinctions are intrinsic, rather than artefacts of the stochastic SF processes. Although we operate on physical scales much larger than the size of a typical giant molecular cloud (GMC; mostly $\gtrsim$300~pc per pixel as in Table~\ref{tab:yes_no} versus $\sim$50~pc, respectively), the stochasticity of the O-type SF might become important in the interarms, since there might not be enough GMCs of sufficient mass to fully populate the high-mass end of the IMF. This would lead to a systematically lower SFR(H$\alpha$) in the interarms as opposed to the arm regions. Moreover, considering that O-stars are very short-lived ($\lesssim$~5~Myr), they might be missing if their original number is low, even though there was a relatively recent SF event. On the other hand, the FUV continuum emission traces OB-stars, and is therefore, less affected in this case.

Previous works have investigated systematically lower H$\alpha$-to-FUV ratios in dwarf galaxies, either from the integrated luminosity \citep{Lee_09} or from fluxes \citep{Meurer_09}. Incorporating stochasticity into the SF modelling showed to produce results aligned with the observations, even for a standard \cite{kroupa_01} IMF \citep{Fumagalli_11}. Given the importance of the stochasticity, it might introduce a larger uncertainty in the fainter interarm regions. To form at least one high-mass star with $M_\star \geq 30\, M_\odot$, the mass of the stellar cluster would vary between a few 10$^4-10^5 \, M_\odot$ \citep[e.g.,][]{calzetti_summer}. For a typical lifespan of an O-star, $\sim$5~Myr, this would correspond to a SFR of 0.002--0.02~M$_\odot$~yr$^{-1}$. In our sample we do encounter pixels where the SFR is below this threshold, both in the arms and interarms, meaning that we are likely to undersample the IMF in those regions. This would lead to an increased scatter in our SFR measurements. However, it is unlikely to have a systematic effect between the arms-interarms and cause the differences in the \dsfr\, distribution in those regions. 

In addition, both H$\alpha$ and FUV emission is susceptible to the contamination from old stellar populations \citep[e.g.,][]{Greggio_90, Hernandez_14, Belfiore_16, Byler_19}. To test whether we are operating in a regime where the emission from old stars dominates over the young ones, we calculated the specific SFR (sSFR) in each pixel, setting a threshold of log(sSFR)~$\leq$~--10.6~yr$^{-1}$, similarly as in \cite{hunt_19}, which corresponds to the turnover in the SFMS in \cite{Salim_07} or \cite{saintonge_22}, for example. We did not find any pixels where this criterion would indicate that the emission mostly comes from old stars.

In summary, we might not be able to fully sample the high-mass end of the ISM in the interarms due to the stochastic nature of the O-type SF in low $\Sigma_\textnormal{SFR(H$\alpha$)}$ environments. Thus, we could be introducing larger uncertainties on the measured \dsfr\, parameter. However, any systematic shifts in the \dsfr\, distribution in the arms and interarms are unlikely. We also did not find any pixels where the H$\alpha$ and FUV emission would be predominantly caused by old stellar populations.

\subsection{Comparison with other works}
In \citetalias{lomaeva_22}, we compared our \dsfr\ to other metrics that have been used to measure recent SFR changes, including ``logSFR79” from \cite{wang_lilly_20} and the $D_{\text{n}}$(4000) break from \cite{kauffmann_03}. The logSFR79 diagnostic is defined using the H$\alpha$ equivalent width, the Lick index of the H$\delta$ absorption, and the 4000~\AA\, break, $D_{\text{n}}$(4000). This makes this index particularly sensitive to the SFR changes that have occurred during the past 5\,Myr, versus 800\,Myr. The $D_{\text{n}}$(4000) break, on the other hand, can indicate stellar populations younger than 1~Gyr when $D_{\text{n}}$(4000) < 1.5 \citep{kauffmann_03}. We found that there were possible scenarios that could explain the changes in the recent SFH as derived by all three metrics as they all probe different timescales. For a more in-depth discussion, see Section~5.2 and Figure~9 in \citetalias{lomaeva_22}.

\begin{figure*} 
	\includegraphics[width=\textwidth]{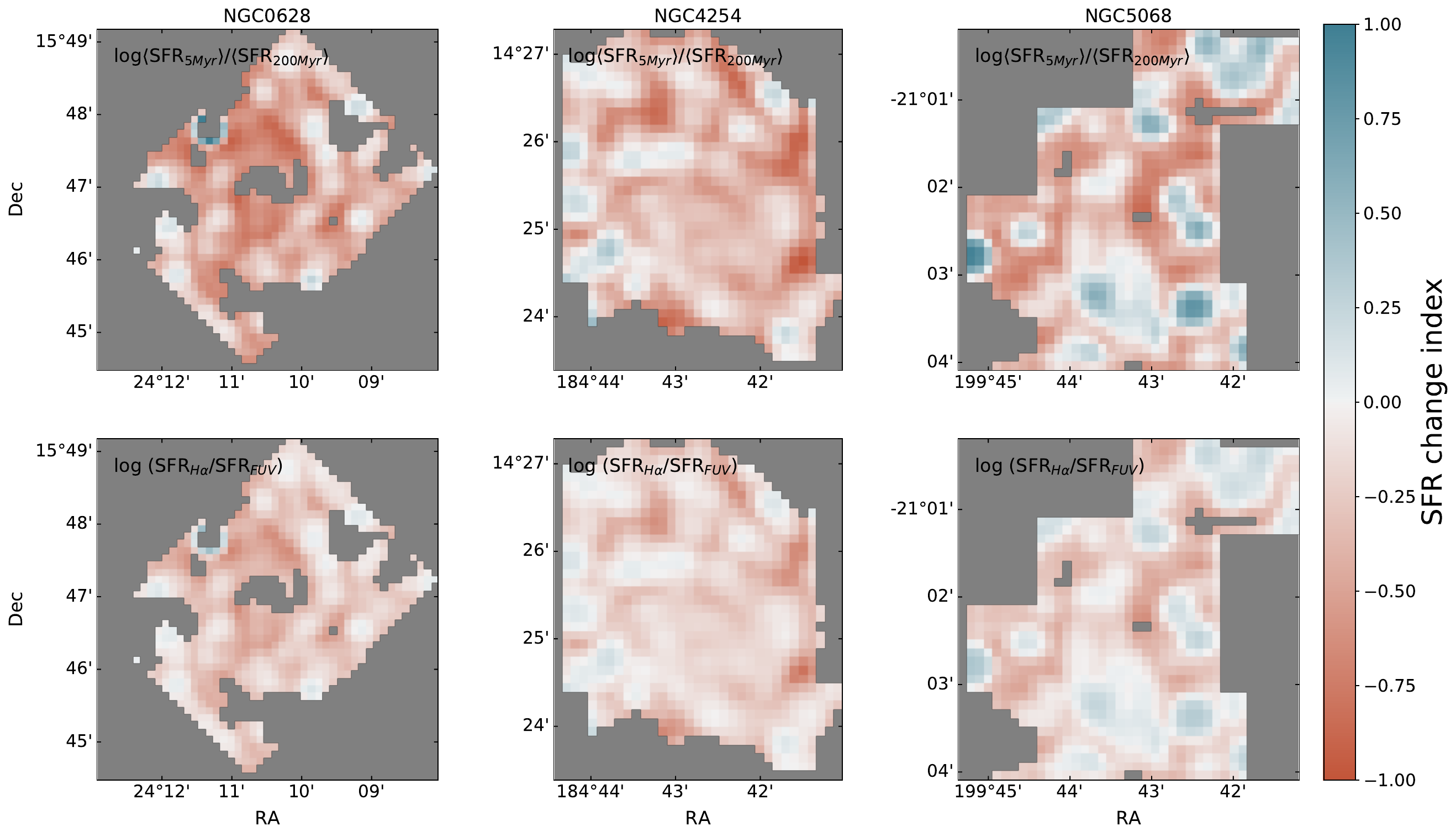}
    \caption{The comparison between \dsfr\ (top row) and log(SFR$_{\textnormal{H}\alpha}$/SFR$_\textnormal{FUV}$) (bottom row) for a selection of galaxies. Both metrics were obtained using attenuation corrected H$\alpha$ and FUV observations.}
    \label{fig:dsfr_sfr_comparison_maps}
\end{figure*}

\begin{figure} 
	\includegraphics[width=\linewidth]{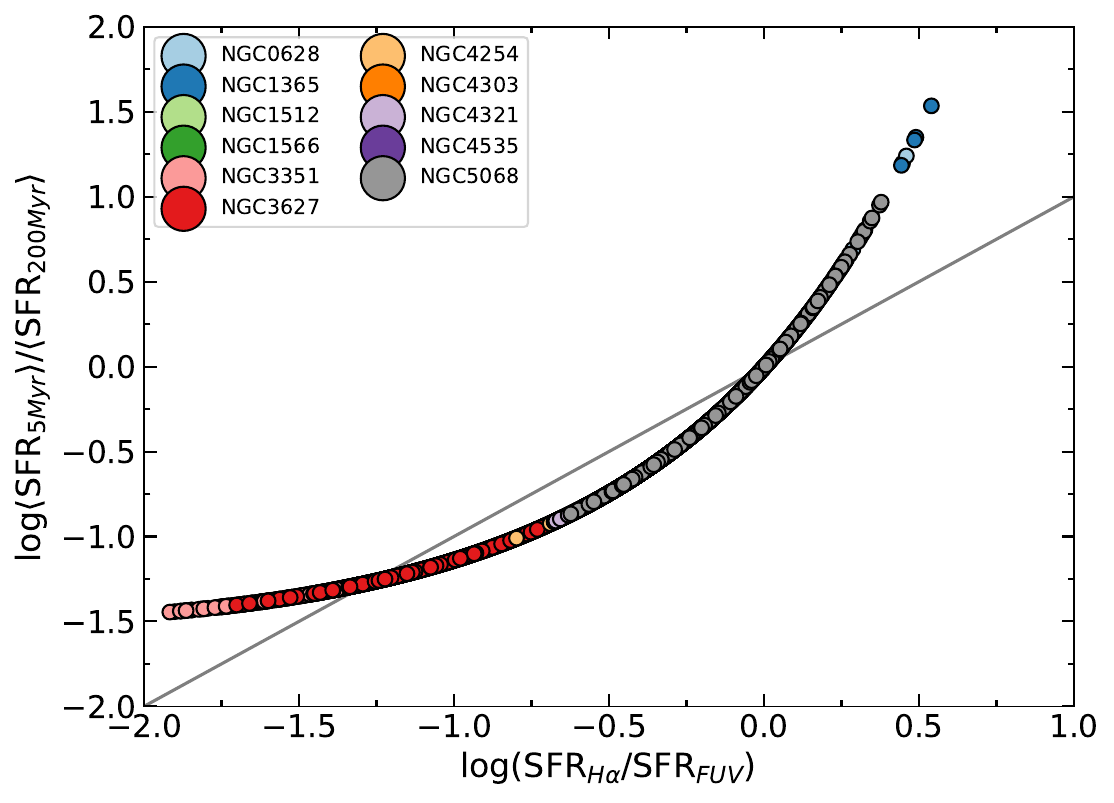}
    \caption{The comparison between the observed log\dsfr\ and log(SFR$_{\textnormal{H}\alpha}$/SFR$_\textnormal{FUV}$) with the solid line showing the one-to-one relation. Again, both metrics were obtained using attenuation corrected H$\alpha$ and FUV observations.}
    \label{fig:dsfr_sfr_comparison_1_to_1}
\end{figure}

There are, however, even more conventional SFR change tracers, such as, log(SFR$_{\textnormal{H}\alpha}$/SFR$_\textnormal{FUV}$), that can reveal similar information about the direction of the recent SF activity. In Figure~\ref{fig:dsfr_sfr_comparison_maps}, we show both log\dsfr\ and log(SFR$_{\textnormal{H}\alpha}$/SFR$_\textnormal{FUV}$) for selected galaxies. Generally, we see a good agreement between both indices, however, \dsfr\ offers a broad parameter range and is therefore better at distinguishing the strength of an increase/decrease in the SF activity. There is another interesting difference: the non-linear relationship between log\dsfr\ and log(SFR$_{\textnormal{H}\alpha}$/SFR$_\textnormal{FUV}$), as demonstrated in Figure~\ref{fig:dsfr_sfr_comparison_1_to_1}. The \dsfr\ metric, therefore, shows that a linear decrease in log(SFR$_{\textnormal{H}\alpha}$/SFR$_\textnormal{FUV}$) does not correspond to a linear decrease in the recent SFR. During the early times of a SF event at high log(SFR$_{\textnormal{H}\alpha}$/SFR$_\textnormal{FUV}$) values, \dsfr\ is very sensitive to the amplitude of the SFR increase, spanning a large dynamical range. When recent SF decreases at low log(SFR$_{\textnormal{H}\alpha}$/SFR$_\textnormal{FUV}$), log\dsfr\ does not drop constantly, but more realistically reaches a plateau. 

\section{Conclusions}\label{sect:conclusions}
In this work, we probed recent SFHs of eleven nearby disk galaxies using the \dsfr\, metric defined in \cite{lomaeva_22}. This parameter is the ratio between the SFR averaged over 5 and 200~Myr as a function of the H$\alpha$--FUV colour and was calibrated in \textsc{CIGALE}. The H$\alpha$--FUV colour was measured using PHANGS-MUSE and GALEX observations, both corrected for the Galactic extinction and internal dust attenuation using NIR and MIR observations. We also examined molecular gas masses with PHANGS-ALMA data. The galaxies in the current sample are star-forming, nearby disks seen mostly face-on. They represent different external (field vs. group/cluster) and internal (AGN vs. inactive; barred, unbarred, and disks without spirals) properties.

\begin{enumerate}
    \item The \dsfr\, diagnostic revealed a diverse behaviour in the observed galaxies. The low global stellar mass NGC~5068 showed the most active recent SF, while the barred NGC~3351 has undergone a rapid drop in the SF across the entire disk visible. NGC~628, 1365, 4254, 4303, 4321, and 4535 host both recent boosts and drops in the SF activity, with the increased recent SFR usually concentrated along the spiral arms where the molecular gas is abundant. NGC~1512, 1566, and 3627 have mostly undergone decreasing SF in the recent past. 
    
    \item We observe no obvious link between a particular internal environment (bar-unbarred), external environment (field vs. cluster/group), feedback (AGN vs. no AGN) and recent SF activity traced by \dsfr. This points at complex gas dynamics, rather than one particular driver of the SF. 
    
    \item The examination of the molecular SFE revealed that it is high in NGC~5068, similarly to \dsfr, as well as in the centre of NGC~1365, which is a barred galaxy with a circumnuclear ring of high SF. According to the K-S test, the SFE distributions in the arms and interarms are intrinsically different, albeit the distribution shapes and median values are rather similar. 
    
    \item Similarly, the \dsfr\, distributions in the arms and interarms are statistically different as well, with the arm population peaking at higher median values. The \dsfr\, distribution in the interarms stretches down to lower values, and these galaxies, therefore, have a more distinct \dsfr\ value between the arms and interarms. Since this behaviour in the arms-interarms is not seen for the SFE, we conclude that this is a matter of recent SFH variations. 
    
    \item In the galaxies with larger differences in the arm-interarm \dsfr\ distributions (about one half of the sample), where \dsfr\, declines more in the interarms than in the arms, we also see higher B/T ratios and earlier morphological types (Sa and Sb vs. Sc). Since the SFE and molecular gas in the arms and interarms appear to be very similar, we conclude that this behaviour of the recent SFHs is driven by other processes that impede recent SF activity in the interarms.

    \item We also discuss potential artefacts that could cause the observed differences in the \dsfr\, in the arms-interarms. First, we examine the effects of variable and constant values of the scaling coefficient $k$ used for correcting for the FUV attenuation. This does not impact the results in a significant manner. We also discuss that we might not be able to fully sample the high-mass end of the IMF in the interarms, which would introduce larger uncertainties and increase the scatter in the SFR measurements. However, we do not expect this to cause any systematic offsets. In addition, we calculated the sSFR and found that the emission from old stellar populations is not sufficient in any of the pixels to contaminate our H$\alpha$ and FUV measurements. 

Generally, this work is an extension of the initial analysis performed on NGC~628 in \cite{lomaeva_22}. It proves that the \dsfr\, metric is a useful tool for probing and comparing recent SFHs between low redshift galaxies on resolved scales. The current sample of eleven galaxies could be extended further as it was merely limited by the availability of the optical and IR data.

\end{enumerate}

\section*{Acknowledgements}

We thank the anonymous referee for the comments that significantly improved the quality of the paper. 
IDL acknowledges support from ERC starting grant \#851622~DustOrigin.This research has made use of the SIMBAD database, operated at CDS, Strasbourg, France. Packages used and not cited in the main text: seaborn \citep{seaborn}; montage is funded by the National Science Foundation under Grant Number ACI-1440620, and was previously funded by the National Aeronautics and Space Administration's Earth Science Technology Office, Computation Technologies Project, under Cooperative Agreement Number NCC5-626 between NASA and the California Institute of Technology.

\section*{Data Availability}

The data used in this article are publicly available through the PHANGS-MUSE, PHANGS-ALMA, the NED and DustPedia databases; any additional data will be shared upon reasonable request to the corresponding author.
 


\bibliographystyle{mnras}
\bibliography{references} 

\bsp	
\label{lastpage}
\end{document}